\documentclass{egpubl}
\usepackage{eurovis2023}

\SpecialIssuePaper         

\CGFStandardLicense

\usepackage[T1]{fontenc}
\usepackage{dfadobe}  

\usepackage{cite}  
\BibtexOrBiblatex
\electronicVersion
\PrintedOrElectronic
\ifpdf \usepackage[pdftex]{graphicx} \pdfcompresslevel=9
\else \usepackage[dvips]{graphicx} \fi

\usepackage{egweblnk}
\usepackage[usenames,svgnames]{xcolor}

\usepackage{subcaption}
\usepackage{tabularx}

\newcommand{\cloud}[1]{\leavevmode{\color{Gray}{\textbf{#1}}}}
\newcommand{\rain}[1]{\leavevmode{\color{SteelBlue}{\textbf{#1}}}}
\newcommand{\lightning}[1]{\leavevmode{\color{Goldenrod}{\textbf{#1}}}}

\title[Designing Defensive Raincloud Plots]%
      {\emph{Teru Teru B\={o}zu}: Defensive Raincloud Plots}

\author[M. Correll]
{\parbox{\textwidth}{\centering Michael Correll$^{1}$\orcid{0000-0001-7902-3907}
        }
        \\
{\parbox{\textwidth}{\centering $^1$ Work completed while at Tableau Research\\
       } 
}
}

\begin{document}

\teaser{
 \centering
  \begin{subfigure}[b]{0.3\textwidth}
         \centering
         \includegraphics[width=\textwidth]{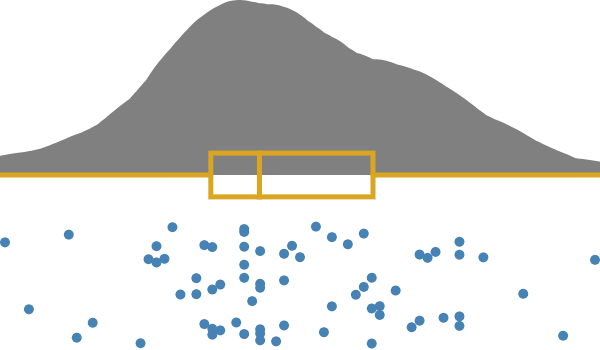}
         \caption{\cloud{Density}-\rain{Jittered}-\lightning{Boxplot}}
         \label{fig:teaserDensity}
  \end{subfigure}
  \begin{subfigure}[b]{0.3\textwidth}
         \centering
         \includegraphics[width=\textwidth]{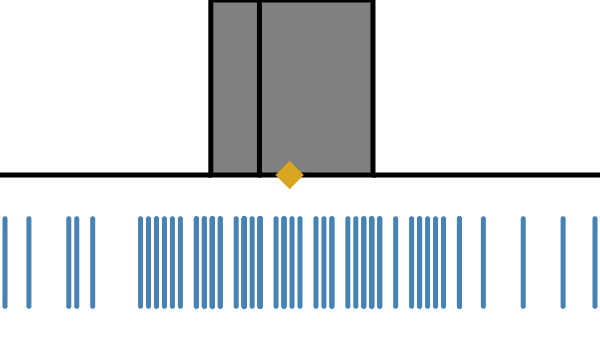}
         \caption{\cloud{Boxplot}-\rain{Strip}-\lightning{Mean Marker}}
         \label{fig:teaserBoxPlot}
   \end{subfigure}
   \begin{subfigure}[b]{0.3\textwidth}
         \centering
         \includegraphics[width=\textwidth]{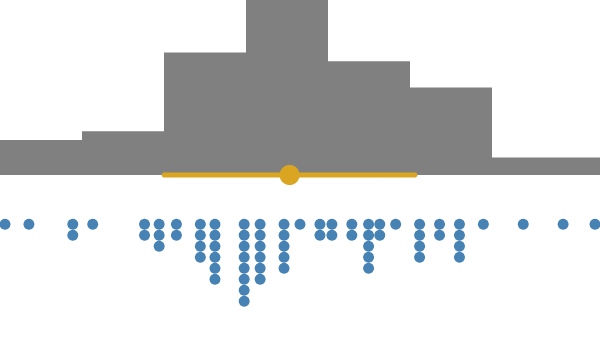}
         \caption{\cloud{Histogram}-\rain{Dot Plot}-\lightning{Mean+Interval}}
         \label{fig:teaserHistogram}
   \end{subfigure}
  \caption{Three examples of raincloud plots showing the distribution of the yearly number of days with precipitation in Seattle, 1948-2017. A raincloud plot is made of three component visualizations: \textbf{{\color{Gray} clouds}} that provide the overall shape of the distribution, \textbf{{\color{SteelBlue} rain}} that plot individual data values, and optional \textbf{{\color{Goldenrod} lightning}} for additional derived or inferential statistics.}
\label{fig:teaser}
}

\maketitle
\begin{abstract}
Univariate visualizations like histograms, rug plots, or box plots provide concise visual summaries of distributions. However, each individual visualization may fail to robustly distinguish important features of a distribution, or provide sufficient information for all of the relevant tasks involved in summarizing univariate data. One solution is to juxtapose or superimpose multiple univariate visualizations in the same chart, as in Allen et al.'s~\cite{allen2019raincloud} ``raincloud plots.'' In this paper I examine the design space of raincloud plots, and, through a series of simulation studies, explore designs where the component visualizations mutually ``defend'' against situations where important distribution features are missed or trivial features are given undue prominence. I suggest a class of ``defensive'' raincloud plot designs that provide good mutual coverage for surfacing distributional features of interest.


\end{abstract}

\section{Introduction}
Prominent examples like Anscombe's quartet~\cite{anscombe1973graphs} or the Datasaurus dozen~\cite{matejka2017datasaurus} show that concise sets of summary statistics may fail to identify distributional features of interest. Visualizations can augment these numerical summaries and provide more context or detail. Yet, simple visualizations like box plots or error bars can similarly fail to surface important distributional features~\cite{choonpradub2005can,potter2006methods}, or, worse, be misinterpreted~\cite{correll2014error}. Even more complex univariate visualizations like density plots and histograms can fail to reliably surface outliers, missing data, or multimodality~\cite{correll2018looks}.

The deficiencies of individual classes of univariate visualizations to support the many, occasionally conflicting tasks involved in univariate distributions have led to a growing number of examples of plots that juxtapose or superimpose \textit{multiple} univariate charts together, where each design is intended to support different sets of tasks relating to the data distribution. The terminology for this class of design varies, for example ``hybrid plots''~\cite{blumenschein2020v,potter2010visualizing}, ``ensemble plots''~\cite{correll2018looks}, or ``RDI plots'' (standing for ``Raw (data), Description and Inference'')~\cite{phillips2017yarrr}. For this paper, I anchor on a specific class of such designs: \textbf{raincloud plots}~\cite{allen2019raincloud} (\autoref{fig:teaser}). As with \textit{teru teru bōzu}, the Japanese good luck charms for warding off rain~\cite{stokes2006charmed}, these rain cloud plots ward off situations where viewers might miss critical information about a data distribution, and promise a sunny day of visual analytics based on trusted and well-understood data.

Raincloud plots juxtapose a ``\cloud{cloud}'' that provides a summary of the overall shape or extent of the distribution (such as a density plot or violin plot) with ``\rain{rain}'' that plots individual data points (such as a jittered dot plot or strip plot). The clouds and rain can be augmented with additional marks indicating medians, normal modes, or inferential statistics~\cite{potter2010visualizing}. I was unable to identify an appropriate extent term for these design elements and so, to extend the weather metaphor, I refer to them as ``\lightning{lightning}.'' The cloud, rain, and lightning, together, present multiple facets of the distribution, at multiple scales, and in the same plot: as per Allen et al.~\cite{allen2019raincloud}, ``the reader has all information needed to assess the data, its distribution, and the appropriateness of any reported statistical tests'' all ``in an appealing and flexible format with minimal redundancy.''

Raincloud plots are still relatively uncommon, earning an entry in Lambrechts' collection of ``\href{https://xeno.graphics/}{xenographics}''--- rare, strange, or unfamiliar forms of visualizations. Even within their niche, there appears to be some disagreement about what elements they must or should contain, or even their general utility. In this paper, I categorize and assess the individual design components of the raincloud plot from the perspective of what Allen et al.~\cite{allen2019raincloud} call ``robustness'' and I call \textbf{defensiveness}: the ability of raincloud plots to consistently and reliably surface distributional features of interest even in the face of potentially adversarial settings. This defensive analysis, in turn, draws on prior work that employs simulation to assess the robustness or reliability of visualization designs~\cite{crisan2021userex,mcnutt2021tables,mcnutt2020surfacing} from the perspective of \textbf{algebraic visualization design}~\cite{kindlmann2014avd} (AVD). That is, two raincloud plots should be visually distinct only to the degree that they represent commensurately distinct underlying distributions and, conversely, two raincloud plots should be visually similar if and only if the underlying distributions encoded by both are commensurately similar.

In this paper, I identify algebraic weaknesses in the underlying univariate visualizations that make up raincloud plots, and how these weaknesses can impact the ability of the resulting plots to act defensively to reveal distributional anomalies. As the result of this analysis, I suggest a set of best practices for raincloud plot design, including avoiding jitter dot plots in favor of more algebraically consistent visualizations of raw values such as strip plots, and towards more compact designs that reinforce the \textit{mutually defensive} aspects of the component plots where the designer accepts some potential deficiencies in individual raincloud components (such as overplotting in dot plots, or oversmoothing in density plots) trusting the other components to ``pick up the slack'' and communicate what might be missed when looking at one component at a time.


\section{Raincloud Plots}
\label{sec:designs}
There are many potential reasons for visualizing a distribution (see Blumenschein et al.~\cite{blumenschein2020v} for a summary of both individual distributional tasks as well as comparative distributional tasks). Beyond direct inspection or comparison, a viewer might wish to ``sanity check''~\cite{correll2018looks} a distribution to assess the quality of the data or verify that the preconditions for inferential statistics or aggregation have been met. Common summaries of visualizations like box plots or confidence intervals provide \textit{some} of the information relevant to such task, but not \textit{all} of it. These simple plots also can produce ambiguities. A boxplot, for instance, can obscure the nature of an underlying distribution, reducing unimodal, bimodal, and skewed distributions to similar or even identical visualizations~\cite{choonpradub2005can,matejka2017datasaurus}. Plotting the raw data directly (with one mark per data value), such as with a dotplot~\cite{wilkinson1999dot}, wheat plot~\cite{few2017wheat}, or strip plot, supports different sets of tasks, but comes with its own drawbacks. For instance, higher-level summary statistics are no longer directly encoded but must be estimated through ensemble visual processes~\cite{szafir2016ensemble}, which may or may not be sufficiently accurate for statistical purposes, as these estimates are likely driven by ``perceptual proxies''~\cite{yuan2019perceptual} rather than explicit statistical calculation. Additionally, as the number of data points becomes large, overplotting or other issues of scale further complicate the process of extracting relevant statistical information and also impact the sheer legibility of the chart.

If high-level visualizations of summary statistics are insufficient to reveal the interior structure of distributions, but low-level visualizations of raw data values are insufficient to reveal higher-level moments or scale to large datasets, a natural solution is to combine the two into a single, mutually-supportive visualization. There have been many proposed solutions in this space. For instance, the canonical violin plot~\cite{hintze1998violin} combines a mirrored density trace with an interior box plot, whereas the bean plot~\cite{kampstra2008beanplot} replaces the box plot with an interior strip plot. Of special note are two ``kitchen sink'' ensemble plots: the summary box plot~\cite{potter2010visualizing} where the mirrored density trace is augmented with redundant density information as well as glyphs for not just means and medians but also higher moments such as skew and kurtosis, and the v-plot~\cite{blumenschein2020v} generalization for custom, dynamic creation of arbitrary ensemble plots for the comparison of distributions (for instance, histograms superimposed on density traces, or violin plots juxtaposed with ``dynamite plots'' of bar charts with error bars).

Of recent prominence in the space of ensemble plots is the raincloud plot, so dubbed by Allen et al.~\cite{allen2019raincloud}. Extending the weather metaphor in the paper, I observe three components in the raincloud plot design, and use a consistent color scheme in the figures and text in the remainder of this paper to differentiate them:

\begin{itemize}
    \item \cloud{Clouds}: high-level summaries of the distribution at a level above the raw data. For instance, density plots (\autoref{fig:teaserDensity}), boxplots (\autoref{fig:teaserBoxPlot}), or histograms (\autoref{fig:teaserHistogram}).
    \item \rain{Rain}: low-level plots of the individual data values. For instance, jittered dot plots (\autoref{fig:teaserDensity}), strip plots (\autoref{fig:teaserBoxPlot}), or Wilkinson-style dot plots~\cite{wilkinson1999dot} (\autoref{fig:teaserHistogram}).
    \item \lightning{Lightning}: glyphs or other visualizations of derived statistics from the distribution. For instance, confidence intervals (\autoref{fig:teaserHistogram}), mean values (\autoref{fig:teaserBoxPlot}), or even glyphs encoding higher moments like skew or kurtosis~\cite{potter2010visualizing}. Note that this term is my own, and there is some ambiguity about whether, e.g., a box plot fulfills either a ``cloud'' or ``lightning'' role, which is perhaps more of a function of visual design than an inherent distinction; in \autoref{fig:teaserDensity}, for instance, the boxplot is in the ``lightning'' role, whereas it takes a ``cloud'' role in \autoref{fig:teaserBoxPlot}.
\end{itemize}

While Allen et al.~\cite{allen2019raincloud} propose several potential combinations of ``\rain{rain}'', ``\cloud{cloud}'', and ``\lightning{lightning}'' in their paper, a dominant design (and the default of the associated \{raincloudplots\} R package) is a raincloud where the cloud is a \cloud{Density} plot (referred to in the paper as a ``split-half violin''), the rain is a \rain{Jittered} dot plot (that is, each data value is plotted as in a scatterplot but where the y-position of the dot is drawn uniformly at random), and the lightning is a \lightning{Boxplot}: \autoref{fig:teaserDensity} is an example of this canonical design. As a back-of-the-envelope assessment of the impact of these design defaults, I conducted a Google Image Search on the term ``raincloud plot'' and collected the first 50 relevant examples (corpus and coding available in supplement). A hand-coded classification of these examples found that \cloud{Density} plots were the overwhelming choice for clouds (48/50 examples): an exception was the use of a \cloud{Boxplot} in the cloud role. \rain{Jittered} dot plots were similarly common as a choice of rain design (40/50), although there were solo occurrences of \rain{Beeswarm} plots, Wilkinson-style \rain{Dot Plots}~\cite{wilkinson1999dot}, and even \rain{Heatmaps} encoding density information. Lightning had the most variability, with \lightning{Boxplots} being the norm (38/50), with the other dominant design being a \lightning{Marker} indicating a central tendency (such as the mean), either with (5/50) or without (2/50) lines for \lightning{Intervals} such as standard error or 95\% confidence intervals (although none of these intervals were labelled in the figure per se; their meaning was often relegated to captions, code snippets, or often unspecified entirely, a common practice~\cite{correll2014error} that can lead to confusion). A minority (4/50) of examples had no lightning at all, but relied on the clouds and rain alone to communicate distributional properties.

Despite the dominance of the ``\cloud{Density}-\rain{Jittered}-\lightning{Boxplot}'' raincloud, I take both the variety of examples in the original Allen et al.~\cite{allen2019raincloud} work and the proliferation of unique examples ``in the wild'' as evidence that there is both mutability in the concept of a raincloud and opportunity to suggest alternative designs. Given the large number of designs for visualizing distributions and the combinatorial explosion introduced by the unification of these designs, the resulting space of potential rainclouds is quite large, and could conceivably be stretched to include a number of prior ensemble plots not otherwise thought of as rainclouds (for instance, a bean plot could be characterized as a ``\cloud{Mirrored Density}-\rain{Strip Plot}-\lightning{Mean Marker}'' raincloud plot). Rather than exhaustively explore this space (if such a list is even possible given the lassitude in definitions around each component: can textual annotations be a form of lightning, for instance?), in the following sections I instead present a ``visualization zoo''~\cite{heer2010tour} of both dominant designs as well as alternative designs of potential empirical interest or prominence in the visualization literature.

\subsection{Designing \cloud{Clouds}}
\label{sec:clouds}
\begin{figure}
    \centering
    \includegraphics[width=0.95\columnwidth]{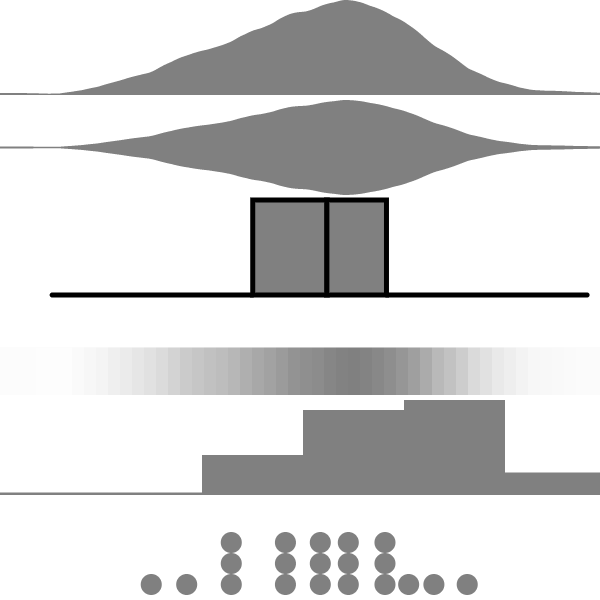}
    \caption{Potential designs for clouds on a sample dataset of 100 values drawn from a Gaussian. From top to bottom: \cloud{Density} plot, mirrored density or \cloud{Violin} plot, ``split-half'' \cloud{Boxplot}, \cloud{Heatmap}, \cloud{Histogram}, and Quantile \cloud{Dot Plot} as in Kay et al.~\cite{kay2016whenish}.}
    \label{fig:clouds}
\end{figure}
As the purpose of a cloud is to provide an overview of the shape of the distribution without plotting individual points, relevant information for cloud designs are kernel density estimates (KDE), bins, or quantile information, visualized through common techniques such as \cloud{Density} plots, \cloud{Histograms}, and \cloud{Boxplots}, respectively. \autoref{fig:clouds} shows a few designs I either observed in the dataset, or suggest as alternatives.

For KDE-driven density plots, prior ensemble designs like violin plots and bean plots often \textit{mirror} the density information. Per Allen et al.~\cite{allen2019raincloud}, ``violin plots mirror the data density in a totally uninteresting/uninformative way, simply repeating the same exact information for the sake of visual aesthetic.'' In addition to redundancy, I also note that mirroring both removes the y-axis baseline for direct comparisons of density and creates negative space between distributions that may or may not reflect genuine distributional features for comparison. Empirically, however, I note that Ibrekk \& Morgan~\cite{ibrekk1987graphical} find this mirroring to not have a significant deleterious effect on performance, and suggest potential benefits in terms of directing attention away from values and into overall shape and area: ``our intent in this [mirrored] display was to try to focus subjects' attention on the \textit{area} between curves.'' Thus, while I argue for the use of non-mirrored (``split violin'') \cloud{Density} plots, and this mirroring is largely inconsequential to the algebraic analysis in this paper, I await further empiricism on this matter.

I note that line and area charts (mirrored or otherwise) are not the only way of communicating density information. I call out two specific designs for having records of potential empirical benefits at distributional tasks. The first are quantile \cloud{Dot Plots}, as introduced in Kay et al.~\cite{kay2016whenish} and examined in follow-on work~\cite{fernandes2018transit,kale2021reasoning}. Quantile dot plots are calculated via a selection of $n$ quantile values and plotted as stacked dots as in a Wilkinson dot plot~\cite{wilkinson1999dot}. The resulting plot approximates the overall shape of the distribution but, unlike in a dot plot of raw values, the number of points shown are capped, affording estimates that rely on counting (for instance, once could count the number of dots below some threshold, and make a frequency-based judgment such as ``in 5 out of $n$ cases, the value is less than the threshold''). While I find quantile dot plots promising for showing the overall shape of a distribution in a bounded-complexity way, I note that, in a raincloud where they are juxtaposed with with a rain component that \textit{does} show raw values, there is potential ambiguity. For instance, a \cloud{Dot Plot}-\rain{Dot Plot} raincloud would likely be confusing to interpret, as there would be two very similar cloud and rain components with very different interpretations (see \autoref{fig:defense} for an example). The second alternative design for distribution shapes are \cloud{Heatmaps} of the KDE. Work by Albers Szafir et al.~\cite{albers2014,szafir2016ensemble} suggests that, while color is less precise as an encoding channel for density information than position or area as used in other designs, they do afford higher precision in extracting \textit{aggregate} statistical information such as regions of high average value or variance. In rainclouds, where the cloud is supplemented by rain and lightning for the purpose of extracting more detailed information, a \cloud{Heatmap} could function as a compact and useful summary. Two-tone~\cite{saito2005two} or horizon charts~\cite{heer2009horizon} represent a potential compromise between traditional density charts and heatmaps, allowing the use of color to draw the eye to salient peaks or valleys of the KDE while maintaining accuracy at extracting individual density values.

A concern shared amongst many of these rain designs is selecting an appropriate kernel and kernel bandwidth (in the case of KDE-driven visualizations) or binning scheme (in the case of histograms). This problem is to some extent a bias-variance tradeoff: if the kernel is too large (or the bins too wide), then fine details of the distribution (like areas of missing data) are lost, but if the kernel is too small (or the bins too narrow), the overall shape of the distribution is lost and spurious visual features like spikes and gaps appear that are a result of overfitting and sampling error rather than genuine modes or gaps in the data~\cite{correll2018looks}. Various rules of thumb for setting these parameters exist (e.g. Sturges' rule~\cite{scott2009sturges}, the Freedman-Diaconis rule~\cite{freedman1981histogram}, Silverman's rule~\cite{sheather2004density}, etc.) based on properties of the underlying data. See Correll et al.~\cite{correll2018looks} for discussion of the algebraic robustness of these parameter settings.

\subsection{Designing \rain{Rain}}
\label{sec:rains}
\begin{figure}
    \centering
    \includegraphics[width=0.95\columnwidth]{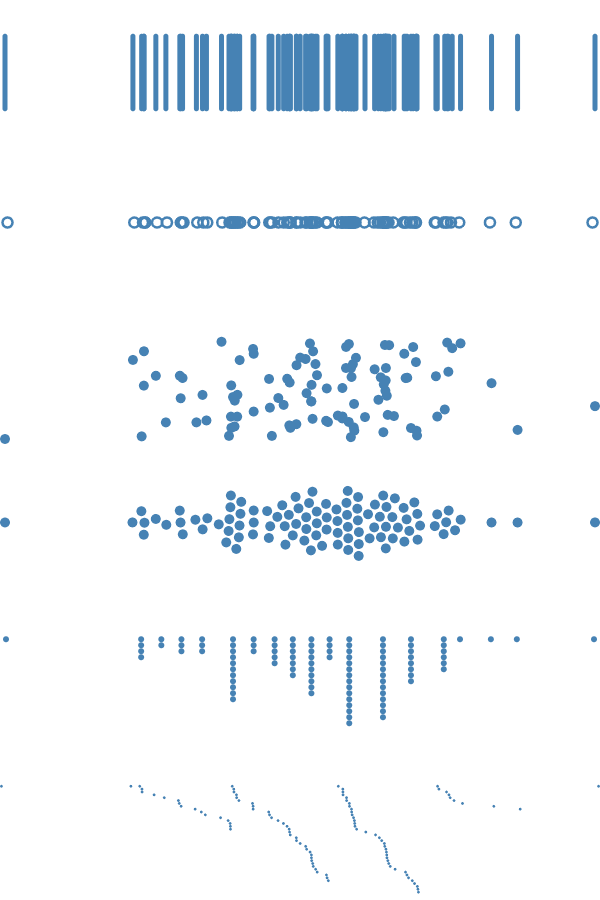}
    \caption{Potential designs for rain on a sample dataset of 100 values drawn from a Gaussian. From top to bottom: \rain{Strip Plot} charts, \rain{Dot Plot}, \rain{Jittered} Dot Pot, \rain{Beeswarm}, Wilkinson \rain{Dot Plot}~\cite{wilkinson1999dot}, and Few \rain{Wheat Plot}~\cite{few2017wheat}}
    \label{fig:rains}
\end{figure}

The rain component of a raincloud plot is meant to plot raw data values directly (see \autoref{fig:rains} for examples). A primary design consideration is therefore the \textit{scalability} and \textit{legibility} of such plots as the number of data values to be plotted increases. For instance, a box plot representing a hundred values can be drawn with the same number of lines as a box plot representing a million, but the same is not true of a rug plot or strip chart. \textit{Overplotting} is perhaps the most well-studied of these scalability issues. I discuss two solutions to overplotting in univariate data that retain the desired outcome of plotting every data value~\cite{few2008solutions}: jittering, and adjusting opacity.

Jittering alters the location of marks by introducing positional offsets. A very simple approach to jittering (and the one encountered most frequently in my dataset of example rainclouds) is to, in the case of marks distributed along an x-axis, to introduce a (meaningless) y-axis and place points randomly within a particular range. In addition to being potentially confusing (as the y-axis does not encode any actual data), this jittering is very disruptive. Few~\cite{few2017wheat} calls jittering ``a carpet bombing that alters the entire landscape (i.e., all of the values) rather than as a surgical strike that targets only those values that are subject to over-plotting and only to the degree that is necessary to resolve the problem.'' Beeswarm plots~\cite{eklund2016beeswarm} are a jittering approach that is less drastic, retaining the secondary axis, but packing marks as closely together as possible. The resulting envelope of the ``swarm'' of points resembles a violin plot. Depending on the data density and the packing algorithm used (the beeswarms in this work are generated via d3's~\cite{bostock2011d3} force-directed simulation functions, for instance), individual marks may end up some distance from their ``actual'' x-axis location. Other jittering approaches make different tradeoffs to respect positional fidelity. For instance, Wilkinson dot plots~\cite{wilkinson1999dot} drop marks in columns. If a mark would be dropped in a column where it would overplot a neighbor, it is instead stacked on top of this mark, and the entire column shifted to represent the mean value of its contents (I note as an aside that the resulting dot plots in the Wilkonson paper are displayed in raincloud-like ensemble plots of dots, histograms, density plots, and strip plots all together). Lastly, Few proposes wheat plots~\cite{few2017wheat}, a sort of hybrid of a dot plot and histogram, where $n$ discrete bins are created, and all the components of each bin are sorted, plotted at their exact x-position, but with a y-position dependent on the index within the bin (the first mark of height $h$ in a particular bin is placed at $y=0$, the next at $y=h$, and the $i$th at $y = hi$, for instance), creating the ``appearance of rows of wheat bending in the wind''~\cite{few2017wheat}. In all of beeswarms, Wilkinson dot plots, and wheat plots, the y-axis now (implicity or otherwise) communicates density information and distributional shape. However, it does so at the cost of compactness: where, in order to fit marks into a given region size chart and so preserve the guarantee of no overplotting, the mark size must be constrained or reduced, eventually producing legibility issues as the size of the data becomes larger or the plot size smaller. \autoref{fig:rains}'s \rain{Wheat Plot} illustrates this issue: with only a small number of bins, each mark has a radius of $\approx2$ pixels in order to accommodate the height of the largest ``stalk'' of wheat; compare to the 5 pixel radius of the \rain{Jittered} dot plot (a more than 6x increase in area).

While positioning can entirely eliminate overplotting, it is also possible to ameliorate its effects by altering how individual marks are rendered. This can be done by reducing the size of the marks, or plotting with an empty rather than filled mark. The latter approach is reflected in the use of an empty circle as the default mark for dot plots and scatterplots in systems with recommendations or smart defaults like VegaLite~\cite{satyanarayan2016vegalite}, Voyager~\cite{wongsuphasawat2017voyager}, and Tableau ShowMe~\cite{mackinlay2007showme}. Reducing the opacity of marks is another approach. Again, as with histogram binning or KDE, there is a tradeoff when altering mark opacity. If the opacity is too low, then individual marks are lost to the background and only the mode(s) are salient. If the opacity is too high, then overplotting means it is difficult or impossible to compare the relative density of dense regions. While there exist automated data-driven methods for setting opacity~\cite{matejka2015dynamic,micallef2017towards}, opacity is often set to a constant by default, and must be interactively adjusted by the chart designer~\cite{matejka2015dynamic}.

One last concern with rain that discourages its use as the only visualization to encode data for which distributional properties are relevant is the difficulty with which the statistics of the overall distribution can be visually estimated from the individual values alone. There are several biases or ``perceptual proxies''~\cite{ondov2020proxies,yuan2019perceptual} potentially present when estimating, for instance, the mean of a group of points. Estimates of the overall shape of the distribution are also potentially biased; for instance, Newburger et al.~\cite{newburger2022umbrella} report a (coincidentally named for the purposes of this paper) ``umbrella effect'' where estimates of distributional shape are adjusted to ``cover'' all observed values, overestimating the variability and estimated density of the distribution in the tails.

\subsection{Designing \lightning{Lightning}}
\label{sec:lightning}
\begin{figure}
    \centering
    \includegraphics[width=0.95\columnwidth]{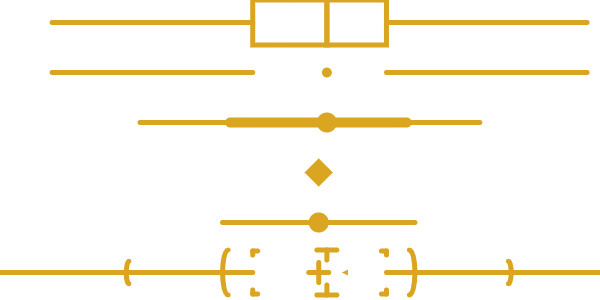}
    \caption{Potential designs for lightning on a sample dataset of 100 values drawn from a Gaussian. From top to bottom: \lightning{Boxplot}, \lightning{Midgap} plot, \lightning{QInterval} showing median, 66\% quantile intervals (thick line), 95\% quantile intervals (thin line), as in the \{ggdist\} R package~\cite{kay2021ggdist}, \lightning{Mean Marker}, \lightning{Mean+Interval} (in this case, one standard deviation), and an adapted form of the Potter et al.~\cite{potter2010visualizing} \lightning{Moment Plot} showing mean (as a +-shaped marker), median (as ``T''-shaped markers), quartiles (as an ``abbreviated'' box), skew (as a triangle), one and two standard deviations (as larger and smaller ``(''-shaped markers), and range (as lines).}
    \label{fig:lightnings}
\end{figure}

``Lightning'' is my grab-bag term for additional marks, annotations, or supplemental charts in a raincloud plot, specifically those meant to directly communicate summary or inferential statistics that are not encoded directly in the cloud or rain components. \autoref{fig:lightnings} presents a set of examples, again based on both observations of raincloud plots as well as suggestions of potentially useful representations from the literature. These statistics can be both descriptive (for instance, means and medians, quartiles or bounds) or inferential (such as confidence intervals, predictive intervals, or indiciations of significant difference from some threshold). Lightning is an opportunity for the designer to directly and saliently encode information that would be difficult or imprecise to extract visually from the other components.

Despite the comparative simplicity in pre-computing and then directly plotting distributional features, there are still important considerations for designers and a multiplicity of final designs. For instance, there are many variations of the canonical \lightning{Boxplot}~\cite{potter2006methods}: the ``whiskers'' can encode the full range of the data, or a scalar multiple of the interquartile range with points outside rendered (or not) with individual glyphs. \lightning{Boxplots} can be further notched to provide guidance on the potential significance of differences between medians, or have variable width to denote sample size, or have continuously varying width driven by a KDE, similar to a violin plot (as in vase plots~\cite{benjamini1988vaseplot}). Lastly, the (relatively large) visual area of the box can introduce a bias in which the widths of the whiskers are underestimated: \lightning{Midgap}~\cite{stock1991midgap} plots are therefore an alternative that still convey identical quartile information while eschewing the central box. A particular variation of note are \lightning{Moment Plots} as described by Potter et al.~\cite{potter2010visualizing}, where box plots are augmented with glyphs representing not just mean and standard deviation but also higher moments such as skew, kurtosis, and tailedness, providing detailed information about the shape and tendencies of a distribution.

I suggest three classes of potential design issues for lightning in the raincloud setting: \textit{ambiguity}, \textit{variability}, and \textit{interpretability}. Lightning is often ambiguous because identical visual designs can have several meanings with different implications for how these designs should be interpreted. The whiskers of the box plot are one example~\cite{potter2006methods} (encoding either range, $1.5\cdot$IQR, or the position of the nearest actual value to $1.5\cdot$IQR). \lightning{Mean+Interval} designs are also ambiguous, because the intervals can represent many different quantities~\cite{correll2014error} (standard deviation, standard error, z- or t-confidence intervals at various $\alpha$ levels, Bayesian predicitive intervals, quantile intervals, etc.), all of which have different implications with respect to interpretation. Lightning is also highly variable, as observed in demonstrations such as Dragicevic's ``dance of the p-values''~\cite{dragicevic2016fair}: data from the same underlying sampling distribution can produce very different inferential statistics and intervals: even with no underlying signal, with enough samples and enough plots, these statistics and their visualization can lead to spurious ``insights''~\cite{zgraggen2018investigating}. Lastly, lightning requires varying levels of statistical familiarity to interpret; while introductory statistics classes may introduce means, medians, and quartiles, lightning visualizing higher moments or more complex derived statistics may be unfamiliar. There is also the task of going from the visual representation of a statistic to a statistical judgment or decision (for instance, using a pair of intervals to estimate the size and reliability of a difference in means between two sample populations). This sort of ``inference by eye''~\cite{cumming2005inference} is non-trivial, and many viewers use ``satisficing'' strategies for this task that can result in biases and inaccuracies in judgments~\cite{kale2021reasoning}

%
%




\begin{table*}
\centering
\resizebox{\textwidth}{!}{
\begin{tabular}{lp{0.65\textwidth}ll}
\hline
Issue                                & Description                                                                                                                                          & Algebraic Failure & Impacted Components \\ \hline
\multicolumn{1}{l|}{Ambiguity}       & Statistical graphics can look visually similar but have distinct underlying meanings and methods of computation.                                     & Hallucinator      & \lightning{Lightning}           \\ \hline
\multicolumn{1}{l|}{Randomness}      & Randomness in layout makes the disambiguation of similar distributions difficult.                                                                    & Hallucinator      & \rain{Rain}                \\ \hline
\multicolumn{1}{l|}{Aliasing}        & Multiple, radically distinct, distributions can have the similar statistical summaries.                                                               & Confuser          & \lightning{Lightning}           \\ \hline
\multicolumn{1}{l|}{Overplotting}    & Comparing relative density is difficult when there are too many overlapping marks.                                                                     & Confuser          & \rain{Rain}                \\ \hline
\multicolumn{1}{l|}{Oversmoothing}   & Kernels that are too large (or bins that are too coarse) can hide important distributional features.                                                 & Jumbler           & \cloud{Cloud}               \\ \hline
\multicolumn{1}{l|}{Discretization}  & Binning produces the impression that the distribution consists of discrete (rather than continuous) values, which may or may not be the case.        & Jumbler           & \cloud{Cloud}, \rain{Rain}         \\ \hline
\multicolumn{1}{l|}{Undersmoothing}  & Kernels that are too small (or bins that are too fine) can produce visual artifacts (spikes and gaps) that are not reliable distributional features. & Misleader         & \cloud{Cloud}               \\ \hline
\multicolumn{1}{l|}{Renormalization} & Adjusting the scales or binning scheme of a plot is visually disruptive, but may not reflect an important change in the data.                         & Misleader         & \cloud{Cloud}, \rain{Rain}        \\
\hline
\end{tabular}
}
\caption{A summary of the raincloud-related AVD violations discussed in this paper.}
\label{tab:avd}
\end{table*}

\section{Defensive Analysis}
\label{sec:avd}

The goal of raincloud plots as specified in Allen et al.~\cite{allen2019raincloud} is ``robustness.'' I interpret this to mean a sort of mutual coverage of the weaknesses of the component parts. For instance, if overplotting in a \rain{Strip Plot} makes it difficult or impossible to assess the relative density of a distribution, a \cloud{Density} plot provides that context. Likewise, if implicitly estimating the mean value from a \rain{Strip Plot} using visual ensemble processes is difficult, a \lightning{Mean Marker} could directly encode that information. I use the term \textit{defensive raincloud} to denote a raincloud in which the subcomponents are mutually supportive in this way. Determining the defensive characteristics of a raincloud, in turn, requires an analysis of the strengths and weaknesses of the subcomponents.

In keeping with prior work examining the robustness of charts~\cite{correll2018looks,crisan2021userex,mcnutt2020surfacing,mcnutt2021tables}, I use the principles of \textit{algebraic visualization design}~\cite{kindlmann2014avd} (AVD) to define potential failures of robustness. While I lay out these failures in more detail in the following sections, in general AVD failures can be thought of mismatches between changes in data and their resulting instantation in a visualization. In an AVD-compliant visualization, changes in data produce \textit{commensurately} important changes in a visualization, and vice versa: trivial changes to data should not result in significant visual changes in the visualization, whereas significant changes in the data should always be visible.  I think this is a reasonable goal in line with the goal of ``robustness'' laid out in Allen et al.~\cite{allen2019raincloud}: a raincloud plot should allow viewers to detect meaningful differences between distributions, while avoiding highlighting trivial differences or creating visual artifacts due to sampling error or poor choices of design parameters.

In the following sections, I present defensive analyses of how raincloud components can produce these sorts of AVD errors. In some cases, it is possible to trivially produce examples of these failures--- in those cases, I rely on visual examples. For others, the scope of the problem is more difficult to assess from single examples; in those cases, I employ simulation to detect the existence of algebraic errors (as in Crisan \& Correll~\cite{crisan2021userex}). There are many families of distributions and a large space of designs and design parameters: I employ simulation here not to fully map this space, but to identify specific issues as they arise across common classes and scales of univariate data.

\autoref{tab:avd} summarizes the AVD issues discussed in this paper. All examples and analyses in this paper are available as an interactive \href{observablehq.com}{Observable} notebook: \href{https://observablehq.com/@mcorrell/raincloud-robustness}{https://observablehq.com/@mcorrell/raincloud-robustness}. The plots themselves are rendered as raster images using \href{https://p5js.org/}{p5.js} to afford pixel-based measures of difference.

\subsection{Hallucinators}
As stated by Kindlmann \& Scheidegger, per AVD~\cite{kindlmann2014avd}:
\begin{quote}
    \textit{The Principle of Representation Invariance (or just Invariance) says that visualizations should be invariant with respect to the choice of data representation: changing the representation should not change the visualization. A visualization failing this principle has a \emph{hallucinator}: a different impression was created (hallucinated, in fact) out of nothing but a different representation of the same data.}
\end{quote}

An example of a hallucinator is a multiclass scatterplot with significant overdraw: in the absence of $\alpha$-blending or other reordering techniques~\cite{chen2018animation}, the order that the various points are drawn can create visually distinct visualizations, despite not representing a genuine change in either the underlying data or visual representation. Given the number of design decisions involved in raincloud plots (not just what sort of components to use, but additional parameters like histogram bins or mark size or opacity), it is not surprising that hallucinators would occur in this setting as well. 

A trivial example I previously discussed is that of \lightning{Interval}-based lightning designs. As shown in \autoref{fig:lightnings}, there are many potential designs that amount, visually, to a point atop a line representing some interval. The choice of \textit{which} interval to visualize can result in lines of radically different visual length (and interpretation) despite sharing the same underlying data and representation. For instance, in my implementation of \lightning{Boxplot} and \lightning{Midgap} plots, I follow the common suggestion of Frigge et al.~\cite{frigge1989some} and have the whiskers denote ranges of $1.5\cdot$IQR from the closest quartile. However, in describing the \lightning{Moment Plot}, Potter et al.~\cite{potter2010visualizing} recommend using whiskers to denote the full range of the data, and so I follow their lead in my implementation. Without this admission, or an inspection of my code, a viewer could be forgiven for thinking that these designs encode different underlying data distributions.

\begin{figure*}
    \centering
    \includegraphics[width=0.85\textwidth]{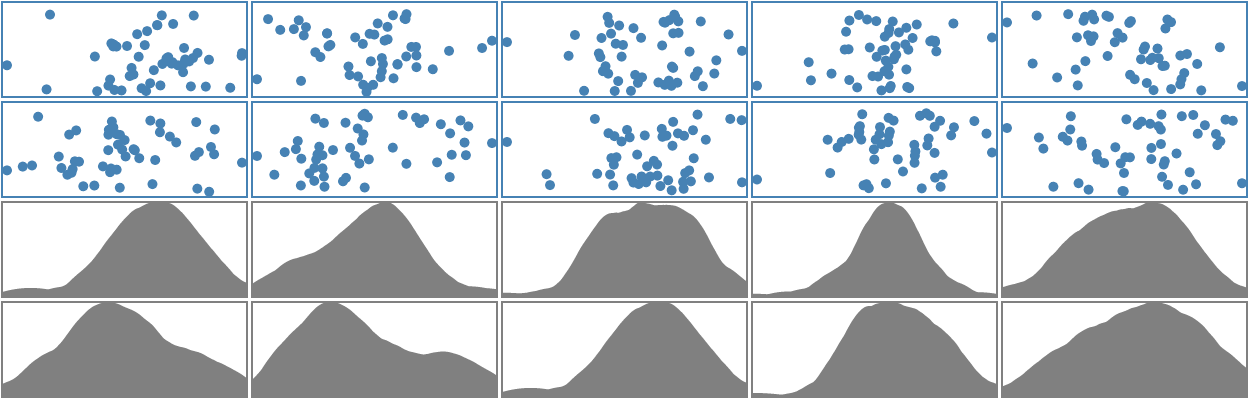}
    \caption{\rain{Jittered} plots \textit{hallucinate} visual differences between plots even with the same data. Two, and exactly two, of each ``lineup''~\cite{wickham2010lineup} of ten charts contain identical data. The other eight are different samples from the the same sampling distribution. Because jittering is randomized on a per-plot basis, it is difficult to disambiguate visual differences caused by data changes versus visual differences caused by jitter. \cloud{Density} plots reveal the identical data (the first and eighth charts in both cases) while still communicating the unimodal roughly bell-shaped nature of all ten samples.}
    \label{fig:jitter-lineup}
\end{figure*}

A more complex hallucinator (and pressing, given the observed dominance of \rain{Jittered} dot plots in rainclouds) arises from the use of random jitter in \rain{Jittered} dot plots. This disruptive ``carpet bombing''~\cite{few2017wheat} of the position channel means that each new rendering of the plot will be visually distinct, even with no changes to the underlying data. This visual distinction makes it very difficult to identify similarities and differences between distributions. The reader is invited to try this exercise for themselves in \autoref{fig:jitter-lineup}. Deterministic layouts of points lack this hallucinatory quality (although are vulnerable to other issues, see below).

\subsection{Confusers}
\begin{figure}[h!]
    \centering
    \includegraphics[width = 0.85\columnwidth]{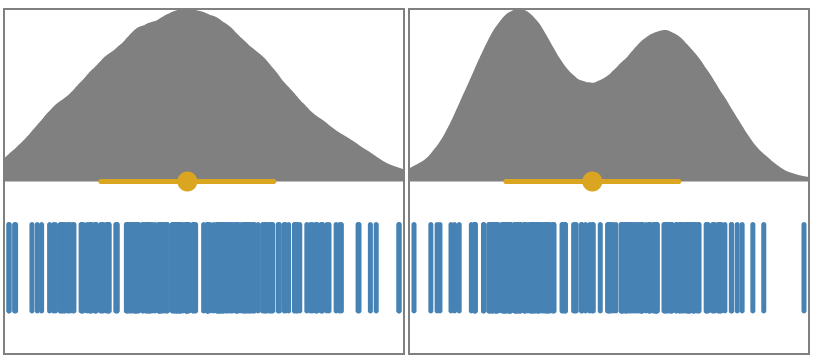}
    \caption{Both overplotting as well as choices of representation can result in \textit{confusion}. Two datasets of 200 points. The data on the left are drawn from a unimodal gaussians, while the data on the right are drawn equally from two unimodal gaussians. The \lightning{Mean+Interval} does not reveal the existence of these two modes. Likewise, overplotting in the \rain{Strip Plot} hides information about modes, as both datasets have significant visual density both near and far from their mode(s). A \cloud{Density} plot is a necessary addition to reveal the hidden shapes.}
    \label{fig:bimodal}
\end{figure}

Again quoting Kindlmann \& Scheidegger on AVD~\cite{kindlmann2014avd}:
\begin{quote}
    \textit{The Principle of Unambiguous Data Depiction (or just Unambiguity) says that visualizations should be unambiguous: changing the underlying data should produce a change in the resulting visualization. Failing this principle, a visualization has \emph{confusers}: changes in the data that are effectively invisible to the viewer of the visualization.}
\end{quote}

There are a number of cases where different data fail to produce visual changes in the resulting univariate visualizations. As with hallucinators above, lightning designs are perhaps the most trivial source of these issues. Choonpradub \& McNeil~\cite{choonpradub2005can}, for instance, present sets of very distinct datasets (such as unimodal Gaussians, bimodal Gaussians, highly skewed data with a significant outlier, etc.) that all have the same five number summaries and so, by definition, all would have an identical \lightning{Boxplot}. Since lightning designs visualize summary statistics, any examples in the long tradition of datasets like Anscombe's quarter~\cite{anscombe1973graphs} that illustrate the weaknesses of these summary statistics can create a confuser. 

However, I note that supplementing lightning with additional plots may not be sufficient to resolve the confuser. \autoref{fig:bimodal} shows an example where the addition of rain is not sufficient to disambiguate data: overplotting and high data density can make distributional shape hard to recover. Likewise, the ``adversarial'' univariate plots in Correll et al.~\cite{correll2018looks} represent attempts to intentionally hide important distributional features by using overplotting, coarse histogram bins, and over-smoothed KDEs.

\subsection{Jumblers \& Misleaders}
\begin{figure}[h]
    \centering
    \includegraphics[width = 0.85\columnwidth]{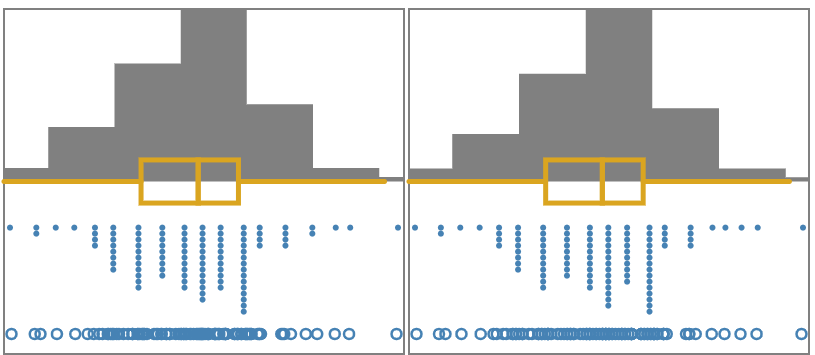}
    \caption{A \cloud{Histogram}-\rain{Dot Plot}-\lightning{Box Plot} that results in \textit{jumbling}: changes to the underlying data do not produce visual differences. On the left are 100 samples from a gaussian ($\mu=0$, $\sigma=20$). On the right, the same data have been converted to integers. The resulting rainclouds are almost entirely identical: the discretization of the data is not visible. The visibility of this change in the rain component, which is intended to reveal these sorts of differences in raw data, is contingent on the size of the marks, the jittering strategy, and the data density. Certain layout choices, such as Wilkinson dot plots, can even further obscure discretization, as they perform a further discretization step of their own.}
    \label{fig:discretized}
\end{figure}

A violation of the last AVD principle produces two classes of errors, again as per Kindlmann \& Scheidegger~\cite{kindlmann2014avd}:

\begin{quote}
    \textit{The Principle of Visual-Data Correspondence (or just Correspondence) says that significant changes in the data should meaningfully correspond with noticeable changes in the visual impression and vice versa. If an important change in data is not clearly manifested in the visualization, it has \emph{jumbled} the data. If a clear and obvious transformation of the visualization corresponds with an unimportant change in the data, the visualization is \emph{misleading}.}
\end{quote}

One concern with the Principle of Correspondence is that both ``significant changes'' in data and ``noticeable changes'' in the reading of a visualization are to some extent dependent on the viewer and their tasks. However, I note that distributional ``sanity checks''~\cite{correll2018looks} often rely on detecting data differences that are relatively small in scale. This means that factors such as overplotting and overaggregating can hide all number of potentially important data changes like missing data~\cite{mcnutt2020surfacing} or changes in internal density. For instance, in \autoref{fig:discretized} the discretization of data is arguably a significant change, but can be difficult or impossible to detect with all three components of a raincloud, contingent on many features of both the design and the data. Visualizations in which these important data changes are difficult to detect are \textit{jumblers}.

\begin{figure}[h!]
    \centering
    \includegraphics[width=0.85\columnwidth]{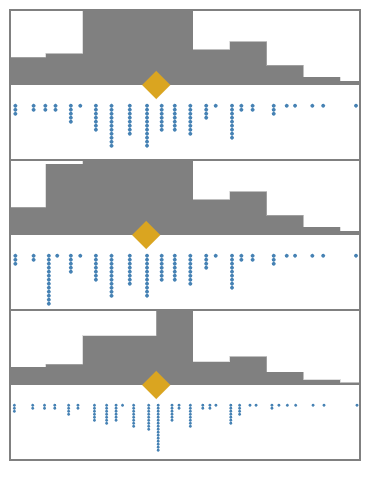}
    \caption{100 points from a gaussian as visualized in a \cloud{Histogram}-\rain{Dot Plot}-\lightning{Mean Marker} raincloud. Adding 10 points to the interior of the distribution ($1.5\cdot\sigma$ from the mean) is not visually disruptive, impacting only a single bin of the \cloud{Histogram} and a single column of the \rain{Dot Plot}, and moving the \lightning{Mean Marker} slightly to the left (middle chart). Adding 10 points to the center of the distribution, however (bottom chart), causes a ``renormalization bias''~\cite{correll2016surprise}: the height of all the bins of the \cloud{Histogram} are impacted, the size of the marks in the \rain{Dot Plot} are decreased, which in turn causes a global re-binning. This is \textit{misleading}, as one would expect visual changes to be commensurate with the importance of the underlying data changes, and these data changes are roughly equivalent (if anything, adding inliers to a sparse region would seem to be more informative than adding to a an existing mode).}
    \label{fig:renormalization}
\end{figure}

The need to bin or aggregate can also create \textit{misleaders}, where visual disruption is not indicative of any particularly relevant or important change in the data, but more an example of hitting arbitrary thresholds in the design parameters. For instance, Correll \& Heer~\cite{correll2016surprise} point to the ``renormalization bias'' in visualizations of density information. Many charts (including all of the raincloud or raincloud component charts in this paper) have \textit{a priori} fixed dimensions and color scales, and so must dynamically adjust scales in order to fit all of the data. Adding a new data point (for instance, to the modal column in a Wilkinson \rain{Dot Plot}) is a relatively minor and unimportant data change (after all, it is already the mode and remains so after the new data point is added), but can cause a visually disruptive renormalization where marks are resized and rebinned to fit in the chart bounds. A \cloud{Heatmap} presents the largest potential for a misleader: adding to the mode causes a renormalization of the entire color scale, impacting almost every pixel in the chart, whereas adding an outlier to an otherwise empty region of the distribution (arguably a more significant data change) only alters pixels in the immediate vicinity of the new data. \autoref{fig:renormalization} shows an example of a potential misleader caused by the renormalization bias. One unique downside of raincloud plots is that these thresholds (and so subsequent visual changes) occur in ways that may not be synced across the three components of the raincloud. For instance, the number of bins in the \cloud{Histograms} shown in this paper are determined via d3's~\cite{bostock2011d3} standard binning function, which employs Sturges' rule~\cite{scott2009sturges} by default. Sturges' rule, in turn, is sensitive only to the number of datapoints. Resizing points in a Wilkinson \cloud{Dot Plot} to fit in a given size is dependent on the maximum number of points in a particular stack of points, rather than the number of bins in total, and so dependent on local density. A visually disruptive rebinning event can therefore occur in a \cloud{Histogram} component of a raincloud plot while leaving all of the other components of the raincloud relatively untouched.

\section{Discussion}
\label{sec:discussion}
The defensive analysis in this papers leads me to a set of conclusions somewhat at odds with the current use and conception of raincloud plots. I condense my findings in the subsections below, with an eye towards influencing the \textit{design} and \textit{use} of this relatively under-explored form, and with an eye towards locations where we (as a community) are in need of \textit{future work} to empirically assess rainclouds and their components.

\subsection{The Unsuitability of the Standard Design}
The \cloud{Density}-\rain{Jittered}-\lightning{Boxplot} raincloud, while not the only variation of the raincloud proposed in Allen et al.~\cite{allen2019raincloud}, is both a common and in some sense emblematic example of the chart type, showing up everywhere from the \href{https://xeno.graphics/raincloud-plot/}{Xenographics entry} for rainclouds to the logo of the \{rainclouds\} \href{https://github.com/RainCloudPlots/RainCloudPlots}{R package}. Yet, I maintain that this design has several key AVD vulnerabilities. 

For one, the \rain{Jittered} dot plot is an almost textbook example of an AVD failure: the random layout of points makes comparison of distributions difficult, and is relatively meaningless despite having large visual impact. I recommend, instead, either a more principled choice of jittering (when the number of points is small enough that individual marks are legible in such designs), or, as a reasonable compromise, a \rain{Strip Plot}. While \rain{Strip Plots} are prone to overplotting, I would still advocate for their use in many scenarios. For one, this overplotting is often acceptable, as overplotting is mainly problematic for hiding density information, and other components of the raincloud can present this information more directly (see \ref{sec:mutual}). Secondly, other rain designs less prone to overplotting can still hide important information about the raw values, and so choosing an alternative does not eliminate this problem. Lastly, the difficulty of estimating means or modes or relative density in overplotted \rain{Strip Plots} might be a ``beneficial difficulty''~\cite{hullman2011beneficial} as viewers will not be tempted to make estimates that may be incorrect.

Another concern with the \cloud{Density}-\rain{Jittered}-\lightning{Boxplot} raincloud is that the standard \lightning{Boxplot} has issues across multiple AVD failure types: it does not represent important distributional quantities in an unambiguous and robust way. I would suggest instead that these marks be more driven by intended tasks and analyses of the viewer, or at the very least ``map''~\cite{fife2021graph} to the actual statistical analyses performed. For instance, if used in a report using Bayesian analyses, a Bayesian credible \lightning{Interval} might be a better choice for lightning. In situations where there is no clear definition of what sort of intervals or summary statistics to expect, a simple \lightning{Marker} for the mean might produce less ambiguity or potential for misuse or misinterpretation.

Beyond avoiding \rain{Jittered} dot plots and \lightning{Boxplots} in rainclouds, I am less dogmatic about the choice of cloud. For discrete data, the visual metaphor of the \cloud{Histogram} might be more appropriate, as there is less temptation to assume that values are continuous than in a \cloud{Density} plot. Similarly,  a \cloud{Density} plot, since the KDE can extend infinitely far in all directions, might suggest the existence of points outside of the observed (or even possible) range of values. Furthermore, there are also cases where there are existing ``semantic bins''~\cite{setlur2022oscar} of meaningful partitions of the data, and so a discretization would be most appropriate or expected for the intended audience. However, these are all relatively narrow and contingent reasons to suggest the use of a \cloud{Histogram}: in general, both \cloud{Histogram} and \cloud{Density} seem to be appropriate choices of cloud. 

\subsection{The Importance of Mutual Defense}
\label{sec:mutual}

\begin{figure}[h!]
     \centering
  \begin{subfigure}[b]{0.85\columnwidth}
         \centering
         \includegraphics[width=\textwidth]{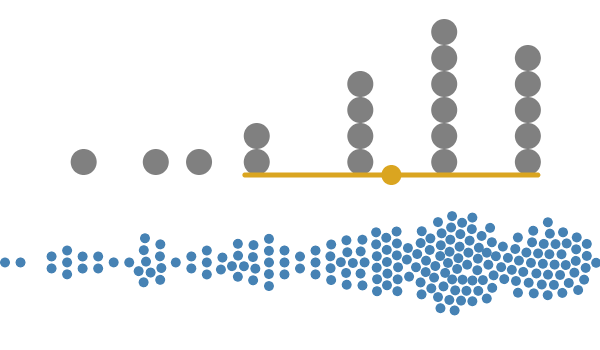}
         \caption{Poor example: \cloud{Dot Plot}-\rain{Beeswarm}-\lightning{Mean+Interval}}
         \label{fig:badRain}
  \end{subfigure}
  
  \begin{subfigure}[b]{0.85\columnwidth}
         \centering
         \includegraphics[width=\textwidth]{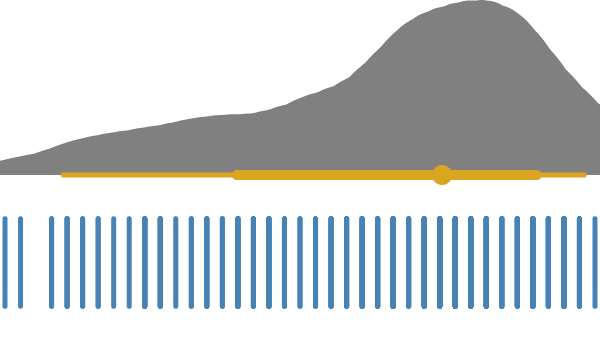}
         \caption{Better example: \cloud{Density}-\rain{Strip}-\lightning{QInterval}}
         \label{fig:goodRain}
   \end{subfigure}
  \caption{In a good raincloud plot, the components are \textit{mutually defensive}. In these examples showing per-country life expectancy in 2012, \ref{fig:badRain} does not provide mutual defense: the \rain{Beeswarm} provides the same information (and more!) about distributional shape as the Quantile \cloud{Dot Plot}. The individual components of \ref{fig:goodRain} have deficiencies, but provides mutual defense: overplotting in the \rain{Strip Plot} makes density hard to recover, but the \cloud{Density} plot can step in to provide it. The \cloud{Density} plot requires the user to extract central tendencies by mentally integrating the KDE, which can be difficult, but the \lightning{QInterval} explicitly encodes this information without the need for a visual estimation.}
\label{fig:defense}
\end{figure}

The existence of AVD issues in a particular raincloud component should not be taken as final proof of their non-utility. The central conceit of a \textit{defensive} raincloud is to permit mutual support and defense between components: while one component might have issues, the raincloud as a whole may not. The lineup in \autoref{fig:jitter-lineup} shows how the introduction of another raincloud component (in this case, the \cloud{Density} plot) can be sufficient to resolve particular classes of AVD errors, in particular \textit{confusers} and \textit{jumblers} where a chart remains the same despite replacement or alteration of the underlying data. Beyond cases where data are intentionally modified to obscure~\cite{matejka2017datasaurus}, or where the scale of the data is such that drastic changes are required to produce visual changes in the estimated density or in the summary statistics, it is difficult (but not impossible!) to conceive of a data change that will not be represented in at least \textit{one} of the components of a raincloud, even if that change is difficult to perceive in the final visualization (see \autoref{fig:defense}). 

However, choices of components should be made with the idea of mutual coverage in mind. Merely having one of each component is not sufficient to guarantee a defence against AVD issues, and not all combinations of elements produce rainclouds that are equally useful (or even coherent: the reader is invited to click the ``I'm feeling lucky'' button in the Observable notebook connected with this paper in order to generate one of 216 possible raincloud designs possible from the components displayed in this work). For instance, \rain{Beeswarms} produce a visual estimation of the density of a distribution; a \cloud{Violin} is therefore not an appropriate choice of cloud, as to some extent the \rain{Beeswarm} has already ``covered'' the visual metaphor of a symmetric representation of estimated density, and perhaps even more thoroughly since the \rain{Beeswarm} affords the reading of individual values. If the \rain{Beeswarm} is legible and sufficiently representative of the density in the underlying distribution, a cloud component may not even be necessary. In that case, it would be beneficial to include a lightning component for indicating summary statistics that would be hard to extract from the \rain{Beeswarm} (e.g., adversarial datasets where estimating means is error prone, as in Ondov et al.~\cite{ondov2020proxies}).

\subsection{The Limitations of Rainclouds}
\label{sec:limitations}
While rainclouds do provide demonstrable benefits for visualizing distributions over their individual components, rainclouds are not panaceas. A viewer must understand the component parts, how they fit together, and be willing to accept the extra complexity (visual and otherwise) that comes with a raincloud plot. Even after these costs are incurred, there is still no guarantee that a raincloud will reveal all distributional features of potential interest (as in \autoref{fig:discretized}). A defensive visualization is not an excuse for the designer to avoid considering the potential tasks or goals a viewer might have, nor an excuse for the viewer to avoid doing due diligence in inspecting their data.

Another limitation is not in the designs of rainclouds themselves, but in the lack of empirical work around these components, either in isolation but (especially) in how they interact. The process of moving between and among different visual representations (and visual metaphors~\cite{ziemkiewicz2008metaphors}) of the same data is an important but understudied assumption undergirding the use of raincloud plots. Similarly, while there is some work in how people ``sanity check''~\cite{correll2018looks} and ``eyeball''~\cite{bartram2022untidy} raw data, and further work in how people build up ensemble statistical pictures from raw data~\cite{szafir2016ensemble} and use that data to make inferences~\cite{kale2021reasoning}, it unclear how the empirical lessons for these disparate lines of work intersect or interact. It is possible, even, that these tasks are diametrically opposed: it could be the case that a viewer looking at visualizations in order to make inferential statistical judgments does not need or want more information about the raw values in the distribution (and in fact this additional information could introduce biases if such judgments are based on inappropriate or inaccurate perceptual proxies). Likewise, a viewer interested in details about individual data values might find estimates of global density or derived summary statistics a distraction. As an example, the sample data used for \autoref{fig:teaser} contains two columns: not just the number of days with precipitation but also the year. A raincloud plot of the year is supremely uninteresting: there is one and only one value for each year. A viewer might wish to confirm this fact, but a raincloud in such a case is excessive when something like a binary assert or column inference, as  in systems like Metareader~\cite{jannah2014metareader}, would complete the job with less ink and less computation.

Another limitation and call for empirical work is with respect to the use of AVD as a tool for assessing rainclouds. While AVD can identify potential violations, a violation is not final proof of non-utility (and, conversely, the lack of violations is not proof of utility). The ultimate utility of a particular design is ultimately determined by how it is used in the real world, with real data and with real viewers. AVD can suggest potential problems or identify areas to explore in more depth, but is not a replacement for an empirical analysis of performance.

\subsection{Conclusion}
Rainclouds remain a promising genre of chart, able to overcome the shortcomings of existing visualizations of distributions by combining the powers and insights from multiple charts into a single view. The relative newness of rainclouds provides us an opportunity to fully explore the space of raincloud designs before the visual genre solidifies, including revisiting old designs for visualizing distributions and inventing new ones. Algebraic visualization design offers us a way of assessing this nascent design space in terms that are aligned with the mission of rainclouds: to visualize distributions without confusion, ambiguity, or waylaid by random chance. Through this algebraic lens, we find that many standard raincloud designs may provide brittle or misinterpretable summaries of distributions, and so should be designed with care, avoiding unnecessary randomness or visual designs sensitive to hyper-parameters not immediately apparent to the viewer. This notion of defensive design can extend to other design problems in visualization as well: the goal of a designer of visualization should be not just to show the data, but to make sure the data have been communicated in a safe, robust, and truthful way.

\bibliographystyle{eg-alpha-doi}  
\bibliography{refs}        


\newpage

\end{document}